\documentclass[a4paper,11pt]{article}
\usepackage[dvips]{color,graphicx}
\usepackage{stmaryrd}

\renewcommand{\b}{\beta}

\newcommand{\s}{\sigma}

\usepackage{graphicx,amssymb,amsmath,bm,latexsym}

\allowdisplaybreaks

\begin{document}
\begin{center}
{\Large\bf Correlators in the supereigenvalue model in the Ramond sector}\vskip .2in
{\large Ying Chen$^{a}$, Rui Wang$^{b}$, Ke Wu$^{a}$,
Wei-Zhong Zhao$^{a}$\footnote{Corresponding author: zhaowz@cnu.edu.cn}} \vskip .2in
$^a${\em School of Mathematical Sciences, Capital Normal University,
Beijing 100048, China} \\
$^b${\em Institute of Applied Mathematics, Academy of Mathematics and Systems Science, 
Chinese Academy of Sciences, Beijing 100190, China}\\

\begin{abstract}
We investigate the supereigenvalue model in the Ramond sector.
We prove that its partition function can be obtained by acting on
elementary functions with exponents of the given operators.
The Virasoro constraints for this supereigenvalue model are presented.
The remarkable property of these bosonic constraint operators is that
they obey the Witt algebra and null 3-algebra.
The compact expression of correlators can be derived
from these Virasoro constraints.
\end{abstract}

\end{center}



\section{Introduction}

Matrix models play important roles in physics and mathematics.
Generally speaking they are quantum field theories where the field is
an $N\times N$ real or complex matrix.
Supereigenvalue models can be regarded as supersymmetric generalizations of matrix models.
They have attracted considerable attention \cite{Zadra}-\cite{Osuga}.
The supereigenvalue model in the Ramond sector is given by \cite{Ciosmak}
\begin{eqnarray}\label{pse}
Z&=&\int d^Nzd^N\theta \Delta_R(z,\theta)^\beta e^{-\frac{\sqrt{\beta}}{\hbar}\sum_{a=1}^{N}V_R(z_a,\theta_{a})},
\end{eqnarray}
where  $d^Nzd^N\theta=\prod_{a=1}^Ndz_ad\theta_a$, $N$ is even, $z_a$ are positive real variables, $\theta_a$
are Grassmann variables, $\Delta_R(z,\theta)$ is the Vandermonde-like determinant,
\begin{eqnarray}
\Delta_R(z,\theta)=\prod_{1\leq a<b\leq N}(z_a-z_b-\frac{1}{2}(z_a+z_b)\frac{\theta_a\theta_b}{\sqrt{z_az_b}}),
\end{eqnarray}
and
\begin{eqnarray}
V_R(z,\theta)=V_B(z)+V_F(z)\frac{\theta}{\sqrt{z}}, \quad V_B(z)=\sum_{k=0}^{\infty}t_kz^k,  \quad V_F(z)=\sum_{k=0}^{\infty}\xi_kz^k,
\end{eqnarray}
$\xi_{k}$ are Grassmann coupling constants,
$V_B(z)$ and $V_F(z)$ are the bosonic and fermionic potentials, respectively.

The various constraints for matrix models have been constructed, such as Virasoro constraints
\cite{Mironov}-\cite{Dijkgraaf},
$W_{1+\infty}$ constraints \cite{Itoyama, Wangr2019}
and Ding-Iohara-Miki constraints \cite{Zenkevich, ZenkevichJHEP}.
They are useful in analyzing the structures of matrix models.
For the partition function (\ref{pse}), it is known that
there are the super Virasoro constraints \cite{Ciosmak}
\begin{eqnarray}\label{svcpse}
L_nZ=\frac{1}{16}\delta_{n,0}Z,\quad\quad  G_nZ=0,\quad\quad n\in \mathbb{N},
\end{eqnarray}
where
\begin{eqnarray}\label{Ln}
L_n&=&\sum_{k=1}^\infty kt_k\frac{\partial}{\partial t_{n+k}}+\sum_{k=0}^\infty(k+\frac{n}{2})\xi_{k}\frac{\partial}{\partial \xi_{k+n}}
+\frac{\hbar^2}{2}\sum_{k=0}^n\frac{\partial}{\partial t_{n-k}}\frac{\partial}
{\partial t_k}+\frac{\hbar^2}{4}n\frac{\partial}{\partial \xi_0}\frac{\partial}{\partial \xi_n} \nonumber\\
&&+\frac{\hbar^2}{2}\sum_{k=1}^{n-1}k\frac{\partial}{\partial \xi_{n-k}}\frac{\partial}
{\partial \xi_k}-\frac{\hbar}{2\sqrt{\beta}}(1-\beta)(n+1)\frac{\partial}{\partial t_n}+\frac{1}{16}\delta_{n,0},
\end{eqnarray}
\begin{eqnarray}\label{Gn}
G_n&=&\sum_{k=1}^\infty kt_k\frac{\partial}{\partial \xi_{n+k}}+\sum_{k=0}^\infty\xi_{k}\frac{\partial}{\partial t_{k+n}}
+\frac{\hbar^2}{2}\frac{\partial}{\partial \xi_0}\frac{\partial}{\partial t_n}+\hbar^2\sum_{k=1}^n\frac{\partial}
{\partial \xi_k}\frac{\partial}{\partial t_{n-k}}\nonumber\\
&&-\frac{\hbar}{\sqrt{\beta}}(1-\beta)(n+\frac{1}{2})\frac{\partial}{\partial \xi_n}.
\end{eqnarray}
The operators (\ref{Ln}) and (\ref{Gn}) obey the super Virasoro algebra
\begin{subequations}\label{witt}
\begin{eqnarray}
&&[L_m,L_n]=(m-n)L_{m+n},\label{witta}\\
&&[L_m,G_n]=\frac{m-2n}{2}G_{m+n},\label{supLG}\\
&&\{G_m,G_n\}=2L_{m+n}-\frac{1}{8}\delta_{m+n,0}\label{supGG}.
\end{eqnarray}
\end{subequations}

Recently a formal supereigenvalue model in the Ramond sector is investigated \cite{Osuga}
\begin{eqnarray}\label{apse}
\breve Z&=&\int \prod_{a=1}^{2N}dz_ad\theta_a \Delta(z,\theta) e^{-\frac{N}{t}\sum_{a=1}^{2N}(z_a^2+V_B(z_a^2)+V_F(z_a^2)\theta_a)},
\end{eqnarray}
where
\begin{eqnarray}
\Delta(z,\theta)=\prod_{1\leq a<b\leq 2N}(z_a^2-z_b^2-\frac{\theta_a\theta_b}{2}(z_a^2+z_b^2)),
\end{eqnarray}
and the bosonic variables $z_a$ are integrated from $-\infty$ to $+\infty$.
To calculate the correlation functions of the model (\ref{apse}),
the recursive formalism has been derived.
It was found that the correlation functions obtained from the
recursion formalism have no poles at the irregular ramification
point due to a supersymmetric correction.

The partition functions of various matrix models can be obtained by acting on
elementary functions with exponents of the given operators,
such as Gaussian Hermitian and complex matrix models and the given $W$ operators
called $W$-representations \cite{Mironov2017}-\cite{Morozov2019}.
For the case of supersymmetric generalizations, to our best knowledge, it has not
been reported so far in the existing literature.
In this letter, we investigate the supereigenvalue model in the Ramond sector and derive its
$W$-representations. We also give the correlators in this matrix model.

\section{Generation of the supereigenvalue model in the Ramond sector by  $\hat W$-operator}
Let us consider the supereigenvalue model in the Ramond sector
\begin{eqnarray}\label{p1}
\bar Z&=&\frac{1}{\Lambda}\int d^Nzd^N\theta \Delta_R(z,\theta)^\beta e^{-\frac{\sqrt{\beta}}{\hbar}
\sum_{a=1}^{N}(V_R(z_a,\theta_{a})+z_a)},
\end{eqnarray}
which can be obtained by taking the shift $t_1\rightarrow t_1+1$ in the bosonic  potential
$V_B(z)$ of (\ref{pse}),  the normalization factor $\Lambda$ is given by
\begin{eqnarray}\label{z0}
\Lambda &=&\int d^Nzd^N\theta \Delta_R(z,\theta)^\beta e^{-\frac{\sqrt{\beta}}{\hbar}
\sum_{a=1}^{N}z_a}.
\end{eqnarray}

We note that the partition function (\ref{p1}) is invariant under
\begin{eqnarray}\label{btran1}
&&z_a\rightarrow z_a+\epsilon\sum_{n=0}^\infty (n+1)t_{n+1}z_a^{n+1}, \quad \theta_a\rightarrow
\theta_a+\epsilon\sum_{n=0}^\infty \frac{n(n+1)}{2}t_{n+1}z_a^n\theta_a,
\end{eqnarray}
with  an infinitesimal bosonic parameter $\epsilon$. It leads to the bosonic loop equation
\begin{eqnarray}\label{loop1}
&&\sum_{n=0}^\infty(n+1)t_{n+1}<-\frac{\sqrt{\beta}}{\hbar}\sum_{a=1}^Nz_a^{n+1}
-\frac{\sqrt{\beta}}{\hbar}\sum_{k=0}^{\infty}(k+\frac{n}{2})\xi_{k}
\sum_{a=1}^Nz_a^{k+n}\frac{\theta_{a}}{\sqrt{z_a}}  \nonumber\\
&&-\frac{\sqrt{\beta}}{\hbar}\sum_{k=1}^{\infty}kt_k\sum_{a=1}^Nz_a^{n+k}
+\frac{\beta}{2}\sum_{k=0}^n\sum_{a,b=1}^Nz_{a}^{n-k}z_b^{k}
+\frac{\beta}{4}n\sum_{a,b=1}^Nz_b^n\frac{\theta_{a}\theta_{b}}{\sqrt{z_az_b}} \nonumber\\
&&+\frac{\beta}{2}\sum_{k=1}^{n-1}\sum_{a,b=1}^Nkz_a^{n-k}z_b^{k}
\frac{\theta_{a}\theta_{b}}{\sqrt{z_az_b}}
+\frac{1-\beta}{2}(n+1)\sum_{a=1}^Nz_a^{n}>=0,
\end{eqnarray}
where the expectation value is taken with respect to the partition function (\ref{p1}).
The loop equation (\ref{loop1}) can be derived by applying the following
differential operators to the partition function (\ref{p1})
\begin{eqnarray}\label{s1}
(\hat{W}_1+\hat{D}_1)\bar Z=0,
\end{eqnarray}
where
\begin{eqnarray}
\hat{D}_1&=&\sum_{k=1}^\infty kt_k\frac{\partial}{\partial t_k}, \nonumber\\
\hat{W}_1&=&\sum_{n,k=1}^{\infty}nkt_nt_k\frac{\partial}{\partial t_{n+k-1}}+\sum_{n=1}^{\infty}\sum_{k=0}^{\infty}n(k+\frac{n-1}{2})t_n\xi_{k}
\frac{\partial}{\partial \xi_{n+k-1}} \nonumber\\
&&+\frac{\hbar^2}{2}\sum_{n=1}^\infty\sum_{k=0}^{n-1}nt_n
\frac{\partial}{\partial t_{k}}\frac{\partial}{\partial t_{n-k-1}}+\frac{\hbar^2}{4}\sum_{n=1}^\infty n(n-1)t_n
\frac{\partial}{\partial \xi_{0}}\frac{\partial}{\partial \xi_{n-1}} \nonumber\\
&&+\frac{\hbar^2}{2}\sum_{n=3}^\infty\sum_{k=1}^{n-2}nkt_n
\frac{\partial}{\partial \xi_{n-k-1}}\frac{\partial}{\partial \xi_{k}}
-\frac{\hbar}{2\sqrt{\beta}}(1-\beta)\sum_{n=1}^\infty n^2t_n
\frac{\partial}{\partial t_{n-1}}.
\end{eqnarray}

The partition function (\ref{p1}) is also invariant under
\begin{eqnarray}\label{btran2}
&&z_a\rightarrow z_a+\epsilon\sum_{n=0}^\infty (n+1)\xi_{n+1}z_a^n\sqrt{z_a}\theta_a, \quad \theta_a\rightarrow
\theta_a-\epsilon\sum_{n=0}^\infty (n+1)\xi_{n+1}z_a^n\sqrt{z_a},
\end{eqnarray}
which leads to another bosonic loop equation
\begin{eqnarray}\label{loop2}
&&\sum_{n=0}^\infty(n+1)\xi_{n+1}<-\frac{\sqrt{\beta}}{\hbar}\sum_{a=1}^Nz_a^{n+1}\frac{\theta_{a}}{\sqrt{z_a}}
-\frac{\sqrt{\beta}}{\hbar}\sum_{k=0}^{\infty}kt_k\sum_{a=1}^Nz_a^{k+n}\frac{\theta_{a}}{\sqrt{z_a}}
-\frac{\sqrt{\beta}}{\hbar}\sum_{k=0}^{\infty}\xi_{k}\sum_{a=1}^Nz_a^{k+n}  \nonumber\\
&&+\frac{\beta}{2}\sum_{a,b=1}^N\frac{\theta_{a}}{\sqrt{z_a}}z_{b}^n
+\beta\sum_{k=1}^n\sum_{a,b=1}^Nz_{a}^k\frac{\theta_{a}}{\sqrt{z_a}}z_b^{n-k}
+(1-\beta)(n+\frac{1}{2})\sum_{a=1}^Nz_a^{n}\frac{\theta_{a}}{\sqrt{z_a}}>=0.
\end{eqnarray}
Similarly, (\ref{loop2}) can be also obtained by applying the following differential operators
to the partition function
\begin{eqnarray}\label{s2}
(\hat{W}_2+\hat{D}_2)\bar Z=0,
\end{eqnarray}
where
\begin{eqnarray}
\hat{D}_2&=&\sum_{k=1}^\infty k\xi_k\frac{\partial}{\partial \xi_k}, \nonumber\\
\hat{W}_2&=&\sum_{n,k=1}^{\infty}nkt_k\xi_n\frac{\partial}{\partial \xi_{n+k-1}}+\sum_{n=1}^{\infty}\sum_{k=0}^{\infty}n\xi_n\xi_{k}
\frac{\partial}{\partial t_{n+k-1}}
+\frac{\hbar^2}{2}\sum_{n=1}^\infty n\xi_n
\frac{\partial}{\partial \xi_{0}}\frac{\partial}{\partial t_{n-1}} \nonumber\\
&&+\hbar^2\sum_{n=2}^\infty\sum_{k=1}^{n-1}n\xi_n
\frac{\partial}{\partial \xi_{k}}\frac{\partial}{\partial t_{n-k-1}}
-\frac{\hbar}{\sqrt{\beta}}(1-\beta)\sum_{n=1}^\infty n(n-\frac{1}{2})\xi_n
\frac{\partial}{\partial \xi_{n-1}}.
\end{eqnarray}

Combining (\ref{s1}) and (\ref{s2}), we have
\begin{eqnarray}\label{sss}
(\hat{W}+\hat{D})\bar Z=0,
\end{eqnarray}
where $\hat{D}=\hat{D}_1+\hat{D}_2$, $\hat{W}=\hat{W}_1+\hat{W}_2$ and
their commutation relation is
\begin{eqnarray}\label{com}
[\hat{D}, \hat{W}]=\hat{W}.
\end{eqnarray}

Since the partition function (\ref{p1}) only depends on even numbers of the fermionic variables,
it can be formally expanded as
\begin{eqnarray}\label{ZC}
\bar Z&=&\sum_{s=0}^{\infty}\bar Z^{(s)}
=e^{-\frac{\sqrt{\beta}}{\hbar}Nt_{0}}\Big[1-\frac{\sqrt{\beta}}{\hbar}C_{k_1}t_{k_1}
+\frac{1}{2!}
(\frac{\sqrt{\beta}}{\hbar})^2C_{k_1, k_2}t_{k_1}t_{k_2}
-\frac{1}{2!}(\frac{\sqrt{\beta}}{\hbar})^2
C^{s_1, s_2}\xi_{s_1}\xi_{s_2}\nonumber\\
&&-\frac{1}{3!}(\frac{\sqrt{\beta}}{\hbar})^3 C_{k_1, k_2, k_3}
t_{k_1}t_{k_2}t_{k_3}+\frac{1}{2!}(\frac{\sqrt{\beta}}{\hbar})^3 C_{k_1}^{s_1, s_2}t_{k_1}\xi_{s_1}\xi_{s_2}+\cdots\Big],
\end{eqnarray}
where
\begin{equation}\label{deg2}
\bar Z^{(s)}=e^{-\frac{\sqrt{\beta}}{\hbar}Nt_0}\Big[\sum_{n=0}^{\infty}
\sum_{\substack{
 m=0}}^{\infty}
\frac{(-1)^{\frac{m(m+1)}{2}}(-\frac{\sqrt{\beta}}{\hbar})^{n+m}}{n!m!}
\sum_{\substack{k_1+\cdots +k_n+\\s_1+\cdots+s_m=s\\k_1,\cdots,k_n\geq 1\\s_1,\cdots,s_m\geq 0}}
C_{k_1, \cdots, k_n}^{s_1, \cdots, s_m}t_{k_1}\cdots t_{k_n}\xi_{s_1}\cdots\xi_{s_m}\Big],
\end{equation}
$m$ is even
and the coefficients $C_{k_1, \cdots, k_n}^{s_1, \cdots, s_m}$ are the correlators defined by
\begin{equation}\label{decor}
C_{k_1, \cdots, k_n}^{s_1, \cdots, s_m}
=\frac{1}{\Lambda}\int d^Nzd^N\theta \Delta_R(z,\theta)^\beta e^{-\frac{\sqrt{\beta}}{\hbar}\sum_{a=1}^{N}z_a}
\sum_{\substack{a_1,\cdots, a_n=1\\b_1,\cdots, b_m=1}}^N
z_{a_1}^{k_1}\cdots z_{a_n}^{k_n}
z_{b_1}^{s_1}\frac{\theta_{b_1}}{\sqrt{z_{b_1}}}\cdots
z_{b_m}^{s_m}\frac{\theta_{b_m}}{\sqrt{z_{b_m}}}.
\end{equation}

For the cases of $m=0$ and $n=0$ in (\ref{decor}), respectively,  we denote
\begin{eqnarray}
C_{k_1, \cdots, k_n}
=\frac{1}{\Lambda}\int d^Nzd^N\theta \Delta_R(z,\theta)^\beta e^{-\frac{\sqrt{\beta}}{\hbar}\sum_{a=1}^{N}z_a}
\sum_{a_1,\cdots, a_n=1}^N z_{a_1}^{k_1}\cdots
z_{a_n}^{k_n},
\end{eqnarray}
and
\begin{eqnarray}
C^{s_1, \cdots, s_m}
=\frac{1}{\Lambda}\int d^Nzd^N\theta \Delta_R(z,\theta)^\beta e^{-\frac{\sqrt{\beta}}{\hbar}\sum_{a=1}^{N}z_a}
\sum_{b_1,\cdots, b_m=1}^Nz_{b_1}^{s_1}\frac{\theta_{b_1}}{\sqrt{z_{b_1}}}\cdots z_{b_m}^{s_m}
\frac{\theta_{b_m}}{\sqrt{z_{b_m}}}.
\end{eqnarray}

Due to the properties of the fermionic variables, we have
\begin{eqnarray}
C_{k_1, \cdots, k_n}^{s_1, \cdots, s_m}=0,\ \   m>N,
\end{eqnarray}
and
\begin{eqnarray}
C_{k_1, \cdots, k_n}^{s_1, \cdots, s_i, \cdots, s_j, \cdots, s_m}=0, \ \  s_i=s_j.
\end{eqnarray}

The operator $\hat{D}$ acting on $\bar Z^{(s)}$ gives
\begin{eqnarray}\label{ds}
\hat{D} \bar Z^{(s)}=s\bar Z^{(s)}.
\end{eqnarray}
By means of (\ref{sss}), (\ref{com}) and (\ref{ds}), we obtain
\begin{eqnarray}\label{wws}
\hat{W}\bar Z^{(s)}=-(s+1)\bar Z^{(s+1)}.
\end{eqnarray}

The partition function (\ref{p1}) is graded by the total $(t,\xi)$-degree.
From (\ref{ds}) and (\ref{wws}), we see that the $\hat{D}$ and $\hat{W}$ are indeed the operators
preserving and increasing the grading, respectively.
In terms of the operator $\hat{W}$, (\ref{ZC}) can be rewritten as
\begin{eqnarray}\label{wopera}
\bar Z&=&\bar Z^{(0)}-\hat{W}\bar Z^{(0)}+\frac{1}{2!}\hat{W}^2\bar Z^{(0)}
-\frac{1}{3!}\hat{W}^3\bar Z^{(0)}+\cdots\nonumber\\
&=&e^{-\hat{W}}\cdot e^{-\frac{\sqrt{\beta}}{\hbar}Nt_0}.
\end{eqnarray}
It indicates that the supereigenvalue model in the Ramond sector
can be obtained by acting on elementary functions with exponents of
the given bosonic operators $\hat W$.

For the $(l+1)$-th power of $\hat{W}$, it can be formally expressed as
\begin{equation}\label{wep}
\hat{W}^{l+1}=\sum_{a,b,c,d=0}^{2(l+1)}
\sum_{\substack{i_1,\cdots,i_a=0\\j_1,\cdots,j_b=0}}^{\infty}
\sum_{\substack{k_1+\cdots+k_c+\\
s_1+\cdots+s_d=\rho\\k_1,\cdots,k_c\geq 1\\s_1,\cdots,s_d\geq 0}}
\hat P^{(k_1, \cdots, k_c|s_1, \cdots, s_d)}_{(i_1, \cdots, i_a|j_1, \cdots, j_b)}t_{k_1}\cdots t_{k_c}\xi_{s_1}\cdots \xi_{s_d}
\frac{\partial}{\partial t_{i_1}}\cdots\frac{\partial}{\partial t_{i_a}}\frac{\partial}{\partial \xi_{j_1}}\cdots\frac{\partial}{\partial \xi_{j_b}},
\end{equation}
where
$\rho=\sum_{\mu=1}^a i_{\mu}+\sum_{\nu=1}^b j_{\nu}+l+1$, the coefficients
$\hat P^{(k_1, \cdots, k_c|s_1, \cdots, s_d)}_{(i_1, \cdots, i_a|j_1, \cdots, j_b)}$ are  polynomials
with respect to $i_{\mu}$, $j_{\nu}$, $k_{\bar\mu}$ and $s_{\bar\nu}$,
$\bar\mu=1,\cdots, c$, $\bar\nu=1,\cdots d$.

Substituting (\ref{wep}) into (\ref{wopera}), comparing the coefficients of
$t_{k_1}\cdots t_{k_n}\xi_{s_1}\cdots\xi_{s_m}$
with $\sum_{\mu=1}^{n}k_{\mu}+\sum_{\nu=1}^ms_{\nu}=l+1$, $k_{\mu}\geq 1$, $s_{\nu}\geq 0$
in (\ref{wopera}) and  (\ref{ZC}), we obtain
\begin{eqnarray}\label{corexp}
&&\frac{(-1)^{l+1}}{(l+1)!}e^{-\frac{\sqrt{\beta}}{\hbar}Nt_0}\sum_{\alpha=1}^{2(l+1)}
\sum_{\sigma_1, \sigma_2}
(-\frac{\sqrt{\beta}}{\hbar}N)^\alpha(-1)^{\tau(\sigma_2(s_1),\cdots ,\sigma_2(s_m))}\hat{P}^{(\sigma_1(k_1),
\cdots, \sigma_1(k_n)|\sigma_2(s_1), \cdots, \sigma_2(s_m))}_{(\underbrace{0, \cdots, 0}_\alpha|\;)}\nonumber\\
&=&
\frac{(-1)^{\frac{m(m+1)}{2}}(-\frac{\sqrt{\beta}}{\hbar})^{n+m}}{n!m!}
e^{-\frac{\sqrt{\beta}}{\hbar}Nt_0}
\sum_{\sigma_1,\sigma_2 }(-1)^{\tau(\sigma_2(s_1),\cdots ,\sigma_2(s_m))}C_{\sigma_1({k_1}), \cdots, \sigma_1({k_n})}^{\sigma_2(s_1), \cdots, \sigma_2(s_m)}\nonumber\\
&=&\frac{(-1)^{\frac{m(m+1)}{2}}(-\frac{\sqrt{\beta}}{\hbar})^{n+m}}{n!m!}
e^{-\frac{\sqrt{\beta}}{\hbar}Nt_0}
\lambda_{(k_1,\cdots ,k_n)}\lambda_{(s_1,\cdots,s_m)}C_{k_1, \cdots, k_n}^{s_1, \cdots, s_m},
\end{eqnarray}
where $\sigma_1$ denotes all the distinct permutations of $(k_1,\cdots ,k_n)$, $\sigma_2$ is all the distinct permutations of  $(s_1,\cdots,s_m)$
and its  inverse number is denoted as $\tau(\sigma_2(s_1),\cdots ,\sigma_2(s_m))$, $\lambda_{(k_1,\cdots ,k_n)}$ and $\lambda_{(s_1,\cdots,s_m)}$
are the numbers of distinct permutations of $(k_1,\cdots ,k_n)$ and $(s_1,\cdots,s_m)$, respectively.

Then we obtain the correlators from (\ref{corexp})
\begin{eqnarray}\label{corr}
C_{k_1,\cdots, k_n}^{s_1,\cdots, s_m}
=\frac{(-1)^{l+1+\frac{m(m+1)}{2}}n!m!(-\frac{\hbar}{\sqrt{\beta}})^{n+m}}
{(l+1)!\lambda_{(k_1,\cdots,k_n)}\lambda_{(s_1,\cdots,s_m)}}\sum_{\alpha=1}^{2(l+1)}
(-\frac{\sqrt{\beta}}{\hbar}N)^\alpha P^{(k_1,\cdots, k_n|s_1,\cdots, s_m)}_{(\underbrace{0,\cdots, 0}_\alpha|\; )},
\end{eqnarray}
where $P^{(k_1, \cdots, k_n|s_1, \cdots, s_m)}_{(\underbrace{0, \cdots, 0}_\alpha|\;)}=\sum_{\sigma_1, \sigma_2}(-1)^{\tau(\sigma_2(s_1),\cdots ,
\sigma_2(s_m))}\hat{P}^{(\sigma_1(k_1), \cdots, \sigma_1(k_n)|\sigma_2(s_1), \cdots, \sigma_2(s_m))}_{(\underbrace{0, \cdots, 0}_\alpha|\;)}$,
$\sum_{\mu=1}^{n}k_{\mu}+\sum_{\nu=1}^ms_{\nu}=l+1$,
$k_{\mu}\geq 1$ and $s_{\nu}\geq 0$.

When particularized to the $m=0$ and $n=0$ cases in (\ref{corr}), respectively,  we have
\begin{eqnarray}\label{corr1}
C_{k_1,\cdots, k_n}
&=&\frac{(-1)^{l+1}n!(-\frac{\hbar}{\sqrt{\beta}})^{n}}
{(l+1)!\lambda_{(k_1,\cdots,k_n)}}\sum_{\alpha=1}^{2(l+1)}
(-\frac{\sqrt{\beta}}{\hbar}N)^\alpha P^{(k_1,\cdots, k_n|\;)}_{(\underbrace{0,\cdots, 0}_\alpha|\; )},\label{corr1}\\
C^{s_1,\cdots, s_m}
&=&\frac{(-1)^{l+1+\frac{m(m+1)}{2}}m!(-\frac{\hbar}{\sqrt{\beta}})^{m}}
{(l+1)!\lambda_{(s_1,\cdots,s_m)}}\sum_{\alpha=1}^{2(l+1)}
(-\frac{\sqrt{\beta}}{\hbar}N)^\alpha P^{(\;|s_1,\cdots, s_m)}_{(\underbrace{0,\cdots, 0}_\alpha|\; )}.\label{corr2}
\end{eqnarray}

For examples, let us list some correlators.

(I) When $l=0$ in (\ref{wep}),  we have
\begin{eqnarray}\label{ex1}
&&P^{(1|\;)}_{(0,0|\;)}=\frac{\hbar^2}{2}, \quad P^{(1|\;)}_{(0|\;)}=-\frac{\hbar}{2\sqrt{\beta}}(1-\beta), \quad P^{(\;|1,0)}_{(0|\;)}=1.
\end{eqnarray}
Substituting (\ref{ex1})
into (\ref{corr1}) and (\ref{corr2}), we obtain
\begin{eqnarray}\label{co1}
C_{1}&=&\frac{1}{\lambda_{(1)}}
\big[-N P^{(1|\;)}_{(0|\;)}+\frac{\sqrt{\beta}}{\hbar}N^2 P^{(1|\;)}_{(0,0|\;)}
 \big]=\frac{\hbar}{2\sqrt{\beta}}N\tilde{N},  \nonumber\\
C^{1,0}&=&-\frac{2\hbar}{\sqrt{\beta}\lambda_{(1,0)}}N
P^{(\;|1,0)}_{(0|\;)} =-\frac{\hbar}{\sqrt{\beta}}N,
\end{eqnarray}
where $\lambda_{(1)}=1$,
$\lambda_{(1,0)}=2$, $\tilde{N}=\beta N+(1-\beta)$.\\

(II) When $l=1$ in (\ref{wep}), we have
\begin{align}\label{ex2}
&P^{(1,1|\;)}_{(0,0,0,0|\;)}=\frac{\hbar^4}{4},
&&P^{(1,1|\;)}_{(0,0,0|\;)}=-\frac{\hbar^3}{2\sqrt{\beta}}(1-\beta),
&&P^{(1,1|\;)}_{(0,0|\;)}=\frac{\hbar^2}{4\beta}(1-\beta)^2
+\frac{\hbar^2}{2}, \nonumber\\
&P^{(2|\;)}_{(0,0,0|\;)}=\hbar^4,
&&P^{(1,1|\;)}_{(0|\;)}=-\frac{\hbar}{2\sqrt{\beta}}(1-\beta), &&P^{(2|\;)}_{(0,0|\;)}=\frac{2\hbar^3}{\sqrt{\beta}}(1-\beta), \nonumber\\
&P^{(1|1,0)}_{(0,0,0|\;)}=\hbar^2,
&&P^{(2|\;)}_{(0|\;)}=\frac{\hbar^2}{\beta}(1-\beta)^2+\frac{\hbar^2}{2},
&&P^{(1|1,0)}_{(0,0|\;)}=-\frac{\hbar}{\sqrt{\beta}}(1-\beta),\nonumber\\
&P^{(\;|2,0)}_{(0,0|\;)}=3\hbar^2,
&&P^{(\;|2,0)}_{(0|\;)}=-\frac{4\sqrt{\beta}}{\hbar}(1-\beta),
&&P^{(1|1,0)}_{(0|\;)}=2.
\end{align}
Substituting (\ref{ex2})
into (\ref{corr}), (\ref{corr1}) and (\ref{corr2}), we obtain
\begin{eqnarray}\label{co2}
C_{2}&=&-\frac{\hbar}{2\sqrt{\beta}\lambda_{(2)}}
\sum_{\alpha=1}^3 (-\frac{\sqrt{\beta}}{\hbar}N)^\alpha P^{(2|\;)}_{(\underbrace{0, \cdots, 0}_\alpha|\;)} \nonumber\\
&=&\frac{\hbar^2}{4\beta}N
(2\tilde{N}^2+\beta),  \nonumber\\
C_{1,1}&=&\frac{\hbar^2}{\beta\lambda_{(1,1)}}
\sum_{\alpha=1}^4 (-\frac{\sqrt{\beta}}{\hbar}N)^\alpha
P^{(1,1|\;)}_{(\underbrace{0, \cdots, 0}_\alpha|\;)} \nonumber\\
&=&\frac{\hbar^2}{4{\beta}}
\tilde{N}N(\tilde{N}N+2), \nonumber\\
C^{2,0}&=&-\frac{\hbar^2}{\beta \lambda_{(2,0)}}
\sum_{\alpha=1}^2 (-\frac{\sqrt{\beta}}{\hbar}N)^\alpha P^{(\;|2,0)}_{(\underbrace{0, \cdots, 0}_\alpha|\;)} \nonumber\\
&=&-\frac{\hbar^2}{2\beta}N(3\tilde{N}+1-\beta),  \nonumber\\
C_{1}^{1,0}&=&\frac{(\frac{\hbar}{\sqrt{\beta}})^3}
{\lambda_{(1)}\lambda_{(1,0)}}
\sum_{\alpha=1}^3 (-\frac{\sqrt{\beta}}{\hbar}N)^\alpha
P^{(1|1,0)}_{(\underbrace{0, \cdots, 0}_\alpha|\;)} \nonumber\\
&=&-\frac{\hbar^2}{2\beta}N(N\tilde{N}+2),
\end{eqnarray}
where
$\lambda_{(2)}=\lambda_{(1,1)}=1$,
$\lambda_{(2,0)}=2$.\\

(III) When $l=2$ in (\ref{wep}), by direct calculations, it is easy to obtain the
precise expression of the 3-th power of $\hat W$. Then we have the final results from (\ref{corr})
\begin{eqnarray}\label{co3}
C_{3}
&=&\frac{1}{8}(\frac{\hbar}{\sqrt{\beta}})^3N[5\tilde{N}^3+(1-\beta)\tilde{N}^2
+10\beta \tilde{N}+3\beta(1-\beta)],  \nonumber\\
C_{1,2}
&=&\frac{1}{8}(\frac{\hbar}{\sqrt{\beta}})^3N(2N\tilde{N}^3
+8\tilde{N}^2+\beta N\tilde{N}+4\beta),  \nonumber\\
C^{2,1}
&=&\frac{1}{4}(\frac{\hbar}{\sqrt{\beta}})^3N
(-2\tilde{N}^2+\beta N\tilde{N}+\beta), \nonumber\\
C^{3,0}
&=&\frac{1}{4}(\frac{\hbar}{\sqrt{\beta}})^3N[-10\tilde{N}^2-9(1-\beta)\tilde{N}
-5\beta-3(1-\beta)^2], \nonumber\\
C_{1,1,1}
&=&\frac{1}{8}(\frac{\hbar}{\sqrt{\beta}})^3N\tilde{N}
(N\tilde{N}+2)(N\tilde{N}+4),  \nonumber\\
C_{1}^{2,0}
&=&\frac{1}{4}(\frac{\hbar}{\sqrt{\beta}})^3N
[-3N\tilde{N}^2-12\tilde{N}-(1-\beta)N\tilde{N}-4(1-\beta)],  \nonumber\\
C_{2}^{1,0}
&=&\frac{1}{4}(\frac{\hbar}{\sqrt{\beta}})^3N
[-2N\tilde{N}^2-13\tilde{N}-3(1-\beta)],  \nonumber\\
C_{1,1}^{1,0}
&=&\frac{1}{4}(\frac{\hbar}{\sqrt{\beta}})^3N
(-N^2\tilde{N}^2-6N\tilde{N}-8).
\end{eqnarray}

\section{Virasoro constraints for the supereigenvalue model in the Ramond sector}
It is known that the partition function (\ref{pse}) is invariant under two pairs of the
changes of integration variables $(z_a\rightarrow z_a+\epsilon z_a^{n+1},
\theta_a\rightarrow \theta_a+\frac{1}{2}\epsilon nz_a^n\theta_a)$ and
$(z_a\rightarrow z_a+z_a^n\sqrt{z_a}\theta_a\delta, \theta_a\rightarrow
\theta_a+z_a^n\sqrt{z_a}\delta)$,  where $\epsilon$ and $\delta$ are the
infinitesimal bosonic and fermionic constants, respectively.
These invariances, respectively, lead to the bosonic and fermionic loop equations
which give the super Virasoro constraints (\ref{svcpse}).
Taking the shift $t_1\rightarrow t_1+1$ in (\ref{svcpse}), we have the super Virasoro constraints for (\ref{p1})
\begin{eqnarray}\label{ssvcpse}
{\bar L}_n\bar Z=\frac{1}{16}\delta_{n,0}\bar Z,\quad\quad  {\bar G}_n\bar Z=0,\quad\quad n\in \mathbb{N}.
\end{eqnarray}
The super Virasoro algebra (\ref{witt})
still holds for the constraint operators ${\bar L}_n$ and ${\bar G}_n$.

From the super Virasoro constraints (\ref{ssvcpse}),  the recursive formulas
for correlators can be obtained. In principle, we can calculate the correlators
step by step from the recursive formulas. However, the compact expression of
correlators (\ref{corr}) can not be derived from them.

Let us introduce the bosonic operators
\begin{eqnarray}\label{Vconst}
\hat L_l=\hat{W}^{l}(\hat{W}+\hat{D}), \quad l\in \mathbb{N}.
\end{eqnarray}
These operators are different from $\bar L_n$.
They obey  not only  the Witt algebra (\ref{witta}),
but also the null Witt $3$-algebra \cite{Zachos}
\begin{eqnarray}\label{n3witt}
[\hat L_{l_1}, \hat L_{l_2}, \hat L_{l_3}]:=\hat L_{l_1}[\hat L_{l_2}, \hat L_{l_3}]-\hat L_{l_2}[\hat L_{l_1}, \hat L_{l_3}]+\hat L_{l_3}[\hat L_{l_1}, \hat L_{l_2}]=0.
\end{eqnarray}

The action of the operators (\ref{Vconst}) on the partition function (\ref{p1}) leads to
the Virasoro constraints
\begin{eqnarray}\label{convc}
\hat L_l\bar Z=0.
\end{eqnarray}
Recently similar Virasoro constraints without the Grassmann
variables have been presented for the Gaussian Hermitian matrix model
and they have been used to derive the correlators of the matrix model \cite{Kang}.

Let us first consider the Virasoro constraints (\ref{convc}) with $l=0$, i.e., (\ref{sss}).
Substituting (\ref{ZC}) into (\ref{sss}), by collecting the coefficients of $t_1^{l}$ and setting to zero,
we obtain
\begin{eqnarray}
C_1=\frac{\hbar}{2\sqrt{\beta}}N\tilde{N},
\end{eqnarray}
and the recursive relations
\begin{eqnarray}\label{wc11}
C_{\underbrace{1, \cdots, 1}_{l+1}}=\frac{\hbar}{2\sqrt{\beta}}
(N\tilde{N}+2l)C_{\underbrace{1, \cdots, 1}_{l}}.
\end{eqnarray}
From (\ref{wc11}), it is easy to obtain
\begin{eqnarray}\label{corr11}
C_{\underbrace{1, \cdots, 1}_{l+1}}=(\frac{\hbar}{2\sqrt{\beta}})^{l+1}
\prod_{j=0}^l(N\tilde{N}+2j).
\end{eqnarray}

We observe that it is difficult to give the precise expression of
$P^{(\overbrace{1,\cdots,1}^{l+1}|\;)}_{(\underbrace{0,\cdots, 0}_\alpha|\;)}$
from $\hat{W}^{l+1}$.
However, by taking $n=l+1$ and $k_1=\cdots=k_n=1$  in (\ref{corr1})
and using (\ref{corr11}), we obtain
\begin{eqnarray}
P^{(\overbrace{1,\cdots,1}^{l+1}|\;)}_{(\underbrace{0,\cdots, 0}_\alpha|\;)}&=&\frac{1}{2^{l+1}}(-\frac{\hbar}{\sqrt{\beta}})^\alpha
\Big[\sum_{\substack{2i+j=\alpha-2\\0\leq i,j\leq l}}
\beta^{i+1}(1-\beta)^j+\sum_{\substack{2i+j=\alpha-1\\0\leq i,j\leq l}}
\beta^{i}(1-\beta)^{j+1}\Big]  \nonumber\\
&&\cdot \sum_{\substack{1\leq r_{1}<r_{2}<\cdots<r_{i+j}\leq {l}\\r_{0}=1}}
\frac{2^{l-(i+j)}\cdot l!}{\prod_{k=0}^{i+j}r_{k}}, \quad \alpha=1,\cdots, 2(l+1).
\end{eqnarray}

Let us collect the coefficients of $t_1^{l}\xi_0\xi_1$ in (\ref{sss}) and set to zero,
we have
\begin{eqnarray}\label{c01}
C^{0,1}=\frac{\hbar}{\sqrt{\beta}}N,
\end{eqnarray}
and the recursive relations
\begin{eqnarray}\label{x0x1}
C^{0,1}_{\underbrace{1,\cdots, 1}_{l}}=\frac{\hbar}{\sqrt{\beta}(l+1)}
[l(l+1+\frac{1}{2}N\tilde{N})C^{0,1}_{\underbrace{1,\cdots, 1}_{l-1}}+NC_{\underbrace{1,\cdots, 1}_{l}}].
\end{eqnarray}
Substituting (\ref{corr11}) into (\ref{x0x1}) we obtain
\begin{eqnarray}\label{corr01}
C^{0,1}_{\underbrace{1,\cdots, 1}_{l}}=2N(\frac{\hbar}{2\sqrt{\beta}})^{l+1}
\prod_{j=1}^l(N\tilde{N}+2j).
\end{eqnarray}

Proceeding the similar procedure for the case of the coefficients
of $t_1^{l}t_2$ in (\ref{sss}), we have
\begin{eqnarray}
C_2=\frac{\hbar}{4\sqrt{\beta}}(2\tilde{N}C_1+\beta C^{0,1})
=\frac{\hbar^2}{4\beta}N(2\tilde{N}^2+\beta),
\end{eqnarray}
and the recursive relations
\begin{eqnarray}\label{re21}
C_{2,\underbrace{1, \cdots, 1}_{l}}&=&\frac{\hbar}{\sqrt{\beta}}\frac{l}{l+2}
(l+3+\frac{1}{2}N\tilde{N})C_{2,\underbrace{1, \cdots, 1}_{l-1}}\nonumber\\
&&+\frac{\hbar}{\sqrt{\beta}(l+2)}(\frac{\beta}{2}C^{0,1}_{\underbrace{1,\cdots, 1}_{l}}
+2\tilde{N}C_{\underbrace{1,\cdots, 1}_{l+1}}).
\end{eqnarray}
Substituting (\ref{corr11}) and (\ref{corr01}) into (\ref{re21}), we obtain
\begin{eqnarray}\label{corr21}
C_{2,\underbrace{1, \cdots, 1}_{l}}=(\frac{\hbar}{2\sqrt{\beta}})^{l+2}N
(2\tilde{N}^2+\beta)
\prod_{j=1}^{l}(N\tilde{N}+2j+2).
\end{eqnarray}

Comparing (\ref{corr01}), (\ref{corr21}) with (\ref{corr}),
we obtain
\begin{eqnarray}
P^{(\overbrace{1,\cdots,1}^{l}|0,1)}_{(\underbrace{0,\cdots, 0}_\alpha|\;)}=P^{(2,\overbrace{1,\cdots,1}^{l-1}|\;)}_{(\underbrace{0,\cdots, 0}_\alpha|\;)}
=0,  \quad \alpha=2(l+1),
\end{eqnarray}
and
\begin{eqnarray}
P^{(\overbrace{1,\cdots,1}^{l}|0,1)}_{(\underbrace{0,\cdots, 0}_\alpha|\;)}&=&
\frac{(-1)^\alpha(l+1)}{2^l}(\frac{\hbar}{\sqrt{\beta}})^{\alpha-1}
\sum_{\substack{2i+j=\alpha-1\\0\leq i,j\leq l}}
\beta^{i}(1-\beta)^{j} \sum_{\substack{1\leq r_{1}<\cdots<r_{i+j}\leq {l}\\r_{0}=1}}
\frac{2^{l-(i+j)}\cdot l!}{\prod_{k=0}^{i+j}r_{k}},   \nonumber\\
P^{(2,\overbrace{1,\cdots,1}^{l-1}|\;)}_{(\underbrace{0,\cdots, 0}_\alpha|\;)}
&=&\frac{l(l+1)}{2^{l+1}}(\frac{-\hbar}{\sqrt{\beta}})^{\alpha+1}
\Big[\sum_{\substack{2i+j=\alpha-3\\0\leq i,j\leq l-1}}
2\beta^{i+2}(1-\beta)^j+\sum_{\substack{2i+j=\alpha-2\\0\leq i,j\leq l-1}}
4\beta^{i+1}(1-\beta)^{j+1}  \nonumber\\
&&+\sum_{\substack{2i+j=\alpha-1\\0\leq i,j\leq l-1}}
(2(1-\beta)^2+\beta)\beta^{i}(1-\beta)^{j}\Big]
\sum_{\substack{2\leq r_{1}<\cdots<r_{i+j}\leq {l}\\r_{0}=1}}
\frac{2^{l-1-(i+j)}\cdot l!}{\prod_{k=0}^{i+j}r_{k}},
\end{eqnarray}
for $\alpha=1,\cdots, 2l+1$.

We have derived the special correlators from (\ref{sss}). It is known that
the compact expression of correlators (\ref{corr}) can not be derived from the
super Virasoro constraints (\ref{ssvcpse}). However,  it should be pointed
out that the special correlators (\ref{corr11}), (\ref{corr01}) and (\ref{corr21})
can be still obtained from (\ref{ssvcpse}).

Let us consider the case of  (\ref{convc}) with $l\neq 0$.
By means of (\ref{sss}) and (\ref{com}), (\ref{convc}) can be rewritten as
\begin{eqnarray}\label{wlz}
\hat{W}^{l+1}\bar Z=(-1)^{l+1}\prod_{j=0}^{l}(\hat{D}-j)\bar Z.
\end{eqnarray}
Substituting (\ref{wep}) into (\ref{wlz}), by collecting the coefficients
of $t_{k_1}\cdots t_{k_n}\xi_{s_1}\cdots\xi_{s_m}$ with $\sum_{\mu=1}^{n}k_{\mu}+\sum_{\nu=1}^ms_{\nu}=l+1$
and setting to zero, we may also derive the correlators (\ref{corr}).

We have achieved the desired correlators from the Virasoro constraints (\ref{convc}).
Unlike the operators $\bar L_n$ in (\ref{ssvcpse}),
the remarkable property of the constraint operators (\ref{Vconst}) is that these bosonic operators
yield the higher algebraic structures.
It should be noted that the closure of the super algebra
does not hold for (\ref{Vconst}) and the fermionic operators $\bar G_n$ in (\ref{ssvcpse}).

\section{Summary}
We have investigated the supereigenvalue model in the Ramond sector
and proved that its partition function can be obtained by acting on
elementary functions with exponents of the $\hat W$ operators.
In terms of the operators $\hat D$ and $\hat W$ preserving and increasing
the grading, respectively, we have constructed the Virasoro constraints
for this supereigenvalue model,
where the constraint operators obey the Witt algebra and null 3-algebra.
The compact expression of correlators (\ref{corr}) can be derived from
these Virasoro constraints. It should be noted that this desired result
can not be derived from the well known super Virasoro constraints (\ref{ssvcpse}).
For the supereigenvalue model in the Neveu-Schwarz sector, whether its partition function
can be expressed in terms of $W$-representation still deserves further study.

We have only constructed the Virasoro constraints for the supereigenvalue model (\ref{p1}).
The remarkable property of these bosonic constraint operators is that they yield the higher
algebraic structures. It is certainly worth to construct the super (Virasoro)
constraints for supereigenvalue models, where the super higher algebraic structures
hold for the bosonic and fermionic constraint operators. It would be interesting to
study further properties of supereigenvalue models from these constraints.

\section *{Acknowledgment}

This work is supported by the National Natural Science Foundation
of China (Nos. 11875194 and 11871350).


\end{document}